# Plate-forme Magicien d'Oz pour l'étude de l'apport des ACAs à l'interaction

Jérôme Simonin — Marius Hategan — Noëlle Carbonell

*Loria – Campus Scientifique*
*BP 239*
*54506 Vandoeuvre-lès-Nancy Cedex*
*France*
*{Jerome.Simonin,Noelle.Carbonell}@loria.fr*

RÉSUMÉ. En vue d'évaluer l'apport des Agents Conversationnels Animés (ACAs) à la communication utilisateur-système, nous avons développé une plate-forme multimodale expérimentale qui met en œuvre le paradigme du Magicien d'Oz pour le recueil et l'analyse des réactions d'utilisateurs à la présence d'un ACA. Cette plate-forme permet de simuler, pour un logiciel quelconque sous Windows, une interface utilisateur réaliste et complexe, dotée éventuellement d'un ACA. Elle permet également de sauvegarder une trace riche des interactions (événements utilisateur et système, copies d'écrans, énoncés et fixations oculaires de l'utilisateur), et de « rejouer » les données enregistrées en vue de leur analyse détaillée. Cette plate-forme est opérationnelle. Nous l'utilisons, dans le cadre d'une étude expérimentale, pour simuler une interface utilisateur multimodale. L'application est une aide en ligne à l'utilisation d'un logiciel grand public (Flash) ; les messages d'aide oraux sont accompagnés, à titre d'illustration, de copies d'écrans de Flash ; ils sont « dits » par un ACA représenté par une tête parlante 3D, développée par FT R&D.

ABSTRACT. In order to evaluate the contribution of Embodied (Animated) Conversational Agents (ECAs) to the effectiveness and usability of human-computer interaction, we developed a software platform meant to collect usage data. This platform, which implements the wizard of Oz paradigm, makes it possible to simulate user interfaces integrating ACAs for any Windows software application. It can also save and "replay" a rich interaction trace including user and system events, screen captures, users' speech and eye fixations. This platform has been used to assess users' subjective judgements and reactions to a multimodal online help system meant to facilitate the use of software for the general public (Flash). The online help system is embodied using a 3D talking head (developed by FT R&D) which "says" oral help messages illustrated with Flash screen copies.

MOTS-CLÉS : Plate-forme logicielle multimodale, agents conversationnels animés, magicien d'Oz, aide en ligne, dialogue homme-machine.

**1. Introduction**

Innovations technologiques et solutions techniques ont permis ces dernières années de réaliser des avancées scientifiques notables en interaction homme-machine concernant, notamment, les modalités d'interaction et l'assistance à l'utilisateur. Ces progrès offrent aux concepteurs d'interfaces utilisateur de nouvelles possibilités pour rendre la communication utilisateur-système, en particulier le dialogue entre les utilisateurs grand public et les services disponibles sur Internet, plus coopératif, convivial et transparent. L'utilisateur peut désormais interagir avec des agents conversationnels animés (ou ACAs), personnages virtuels doués de capacités d'expression verbale et non verbale, mouvements, gestes et expressions faciales. Ces agents peuvent être représentés graphiquement sous la forme de clones humains 3D.

Ainsi, de nouvelles formes d'interaction multimodale sont envisageables grâce aux ACAs, en particulier pour la conception de logiciels destinés au grand public. Toutefois, hormis les travaux de recherche visant à évaluer l'apport des têtes parlantes à l'intelligibilité des messages oraux pour les mal-entendants [MAS 98], les études portant sur l'évaluation de l'apport des ACAs « parlants » à la communication entre les utilisateurs et le système sont rares. En outre, la majorité d'entre elles portent sur des aspects spécifiques de l'interaction.

A titre d'exemple, [BUI 04] étudie l'influence de trois stratégies différentes de désignation multimodale mises en œuvre par un ACA, sur l'efficacité de l'interaction (i.e., rapidité d'exécution d'instructions) pour diverses catégories d'utilisateurs. [WAN 05] montre que l'utilisation de formules de politesse par un ACA tuteur a une influence positive, à la fois sur la motivation des apprenants et sur leur assimilation des concepts difficiles, par rapport à une stratégie de communication classique, plus neutre, de la part du système. Plus généralement, [ROS 03] constate le faible volume des travaux publiés et des résultats scientifiques obtenus sur les nombreuses questions de recherche d'ordre ergonomique que soulève l'introduction d'ACAs en interaction homme-machine. Ce constat n'est pas surprenant compte tenu de la jeunesse de ce domaine de recherche.

Quant à l'évaluation des ACAs eux-mêmes, elle fait intervenir essentiellement des techniques de recueil des jugements subjectifs d'utilisateurs potentiels, questionnaires verbaux, interviews, 'focus groups'. Jusqu'à présent, comme l'affirme [RUT 02], peu d'études ont eu recours à l'observation et à des analyses d'usage dans des situations réelles ou réalistes. L'utilisation de mesures objectives, indicateurs physiologiques ou suivi du regard, est également rare, malgré les recommandations formulées dans [MOR 03] et [MA 05].

Compte tenu des observations qui se dégagent de la revue des travaux de recherche récents sur l'évaluation ergonomique de l'apport des ACAs à la communication utilisateur-système et, en particulier, des manques observés, nous avons décidé de recueillir les réactions d'utilisateurs à la présence d'un ACA. Nous avons donc conçu et réalisé une plate-forme logicielle permettant la simulation d'interfaces utilisateur intégrant un ACA. Cette plate-forme générique, qui utilise le paradigme du Magicien d'Oz en tant que technique de prototypage rapide, fournit une assistance logicielle évoluée au compère pour la simulation réaliste d'interfaces multimodales utilisateur où le système est représenté par un ACA doué de capacités d'expression verbale et non verbale. Elle permet également de sauvegarder une trace des interactions sous forme numérique plutôt qu'analogique, donc plus riche qu'un enregistrement vidéo et utilisable directement pour des traitements automatiques. Elle offre enfin la possibilité de « rejouer » et d'annoter les données multimodales enregistrées. La plate-forme est opérationnelle : nous l'utilisons actuellement pour étudier le comportement et les jugements subjectifs d'utilisateurs d'un système d'aide en ligne multimodal à l'utilisation d'un logiciel grand public (Flash). Pour simuler les réactions du système d'aide aux requêtes d'utilisateurs novices sur le fonctionnement de Flash, il suffit au compère d'activer l'un des message d'aide multimodaux prédéfinis, sauvegardés sur son poste de travail. La composante graphique du message s'affiche en temps réel sur l'écran du sujet dans une fenêtre dédiée ; elle comprend une ou plusieurs copies d'écran de Flash et la représentation graphique animée de l'ACA, une tête parlante au cas particulier, qui énonce la composante orale du message en accompagnant l'énoncé d'expressions faciales et de mouvements de tête appropriés.

Nous décrivons dans la suite de cet article l'architecture fonctionnelle globale de cette plate-forme, ainsi que les choix de conception et de mise en œuvre qui ont présidé à sa réalisation. La description est illustrée d'exemples empruntés à l'étude expérimentale évoquée ci-dessus.

## 2. Architecture

La plate-forme logicielle dont l'architecture fonctionnelle est présentée dans la figure 1 a été développée sous Windows. Elle vise une triple finalité fonctionnelle :
– Enregistrer la trace des interactions de l'utilisateur courant d'un logiciel grand public quelconque ou d'une application propriétaire ;
– Mettre en œuvre la technique du magicien d'Oz pour la simulation rapide, réaliste et robuste d'interfaces utilisateur multimodales intégrant un ACA ;
– Assurer le « rejeu » des interactions enregistrées en vue de leur analyse.

### 2.1. Enregistrement des interactions

Pour comprendre et modéliser le comportement d'utilisateurs lors de leur interaction avec un logiciel, il est souvent nécessaire de disposer d'une trace détaillée de leurs actions et des effets de celles-ci sur le logiciel. La solution classique consiste à réaliser un enregistrement vidéo de l'utilisateur et de l'écran de son poste de travail. Mais l'inconvénient majeur de cette solution est d'imposer un pré-traitement « manuel » des données. Un enregistrement numérique permet en revanche un traitement automatique ou semi-automatique des données brutes.

Notre plate-forme permet l'enregistrement daté des informations suivantes :
– Les événements déclenchés par les actions de l'utilisateur sur l'interface du logiciel. Ils sont de deux types : « système » (déplacement de fenêtres, accès à un menu, ouverture d'une boîte de dialogue, etc.) et « utilisateur » (mouvement de la souris, clic souris ou frappe d'une touche du clavier).
– Les cordonnées en X et Y de la position du curseur souris sur l'écran.
– Les cordonnées en X et Y des positions successives du regard de l'utilisateur sur l'affichage. L'oculomètre utilisé est le modèle 501 d'ASL, la fréquence d'échantillonnage est de 60 Hz.
– Les énoncés éventuels de l'utilisateur, sous la forme de fichiers WAV.
– Des copies d'écran, sous la forme d'images JPEG générées à chaque événement utilisateur ou système.

A noter qu'en l'absence d'événement (i.e., d'action de l'utilisateur ou du système sur l'interface) des copies d'écran sont réalisées à une fréquence paramétrable (i.e., par génération d'un événement « automatique »). Les coordonnées des positions successives du regard sont datées lors de leur arrivée sur le port série. L'événement correspondant au lancement de l'acquisition d'un énoncé permet de dater le début de celui-ci. L'heure système est utilisée pour toutes les datations.

L'application qui réalise ces enregistrements a été développée en Java 1.5, à l'aide de l'outil de développement Eclipse 3.2. Elle permet, si l'on utilise des PC standard, d'enregistrer et d'envoyer au compère, via le réseau local, entre 3 et 5 événements par seconde.

**2.2. Intégration et mise en œuvre de l'ACA**

L'ACA utilisé est de technologie ActiveX, donc conçu pour être intégré à une page HTML. Piloter son animation depuis une application Java présente certaines difficultés d'intégration, mais offre une flexibilité de mise en œuvre supérieure. Les développements réalisés permettent de piloter l'animation par une méthode Java à laquelle on fournit en paramètre le fichier SMIL décrivant l'animation. Par exemple, pour l'étude expérimentale en cours, nous avons défini comme suit les 300 et quelques messages d'aide multimodaux nécessaires. Pour chaque énoncé que devait prononcer l'ACA, nous avons créé un fichier SMIL qui utilise deux fichiers : l'un décrit l'énoncé (fichier texte ou WAV), l'autre l'animation de l'ACA, ses expressions faciales et ses mouvements de tête au cas particulier.

**2.3. Magicien d'Oz, poste « compère »**

Notre implémentation de la technique du magicien d'Oz autorise, pour l'instant, l'intervention d'un seul compère, ce qui peut représenter une contrainte gênante pour la simulation d'interfaces utilisateur complexes ; voir, par exemple, [ROB 00] où la simulation d'une interface multimodale (parole + geste) fait intervenir deux compères, l'un pour l'exécution des commandes multimodales de l'utilisateur sur l'application graphique, l'autre pour l'activation de messages oraux pré-enregistrés.

Notre plate-forme utilise le réseau local pour assurer les échanges d'information à distance entre le compère et le sujet. Le protocole utilisé pour le transport des copies d'écran est UDP, pour des raisons de performance essentiellement étant donné la fréquence assez élevée des événements déclencheurs de copies d'écran. Au cas particulier, UDP offre une fiabilité suffisante, compte tenu de la nature des informations transportées[1]. En revanche, les requêtes de l'utilisateur sont transmises au compère par « sockets », liaison sans perte, car ce sont des séquences d'événements utilisateur. Enfin, pour éviter d'accroître inutilement la charge du réseau, la base de données des messages multimodaux est installée sur le poste du sujet. Le miroir de cette base de données sur le poste du compère ne contient que les commandes nécessaires à l'activation des messages, ce qui limite le volume des données transitant sur le réseau.

Pour suivre l'activité du sujet dans des situations d'interaction courantes, le compère dispose d'une trace suffisante de l'évolution des affichages sur le poste de celui-ci, puisqu'elle comprend, sur un PC standard, 3 à 5 copies d'écran par seconde. Il peut également, le cas échéant, recevoir les énoncés du sujet. En revanche, pour éviter de surcharger son écran, la trace des mouvements oculaires du sujet ne lui est pas transmise. A noter que cette fonctionnalité est facile à mettre en œuvre puisque les données oculométriques datées sont sauvegardées ; elle peut s'avérer nécessaire, par exemple pour simuler une interface où le regard remplace la souris en tant que modalité de pointage et de sélection des objets affichés.

La plate-forme étant ouverte, il est possible d'offrir au compère une assistance logicielle pour la réalisation de certaines des tâches impliquées dans la simulation des réactions de l'interface utilisateur du logiciel considéré. L'intégration à la plate-forme d'outils spécifiques d'aide à l'activité du compère est très facile et rapide à réaliser. Par exemple, dans le contexte de l'étude expérimentale en cours, nous avons développé et intégré à cette plate-forme un composant logiciel de filtrage des messages d'aide disponibles (plus de 300). Ce composant, à partir de l'analyse d'une requête du sujet au système d'aide simulé, propose au compère un choix limité de messages ; il ne reste plus à celui-ci qu'à sélectionner à la souris celui d'entre eux qui lui semble le plus pertinent pour satisfaire la requête du sujet. Ce composant, réutilisable pour la simulation d'aides en ligne à la plupart des logiciels destinés au grand public, permet de réduire la charge de travail du compère et d'obtenir des temps de réponse du système d'aide simulé qui soient réalistes, dans la mesure où la recherche « manuelle » d'un item donné dans un ensemble de plusieurs centaines d'informations est beaucoup plus lente qu'une recherche automatique, même si l'ensemble est structuré en base de données ; elle est également moins fiable. Doter le compère d'une assistance logicielle permet également de stabiliser son comportement, d'éviter que celui-ci ne dérive ou fluctue au cours du temps, sous l'influence de phénomènes d'apprentissage ou sous l'effet de la fatigue.

---

[1] Ce protocole est utilisé également, le cas échéant, pour la transmission des énoncés des sujets vers le compère.

**2.4. « Rejeu » des données sauvegardées**

La plate-forme assure le « rejeu » synchronisé de la trace des interactions enregistrées (voir la description présentée dans la section 2.1). Les fixations oculaires calculées à partir des échantillons sauvegardés apparaissent superposées aux copies d'écran, ce qui permet d'identifier les éléments de l'affichage qui mobilisent l'attention visuelle de l'utilisateur au cours de l'interaction.

Nous envisageons d'enrichir ce composant qui ne permet actuellement que la visualisation interactive de la trace des interactions, en offrant aux chercheurs des outils évolués d'annotation « manuelle » ou semi-automatique, susceptibles de leur faciliter significativement l'analyse détaillée de ces traces, impossible à réaliser dans des temps raisonnables de façon entièrement « manuelle », en raison de leur richesse.

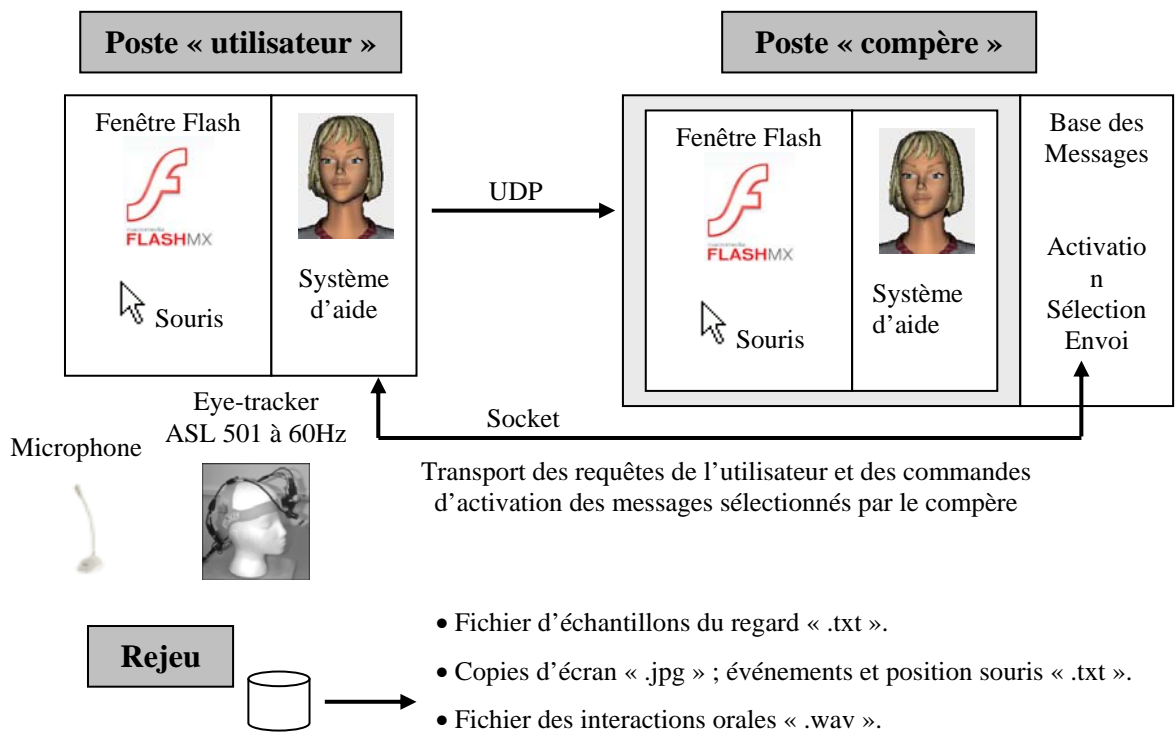

**Figure 1.** *Architecture logicielle globale de la plate-forme.*

## 3. Exemple d'utilisation

Dans le contexte de l'étude expérimentale en cours, nous utilisons l'ensemble des fonctions de la plate-forme : simulation partielle de l'interface utilisateur par la technique du magicien d'Oz, enregistrement et « rejeu » de la trace des interactions de l'utilisateur avec une application Windows et, simultanément, avec une application intégrant un ACA, que nous avons développée. L'étude porte sur l'évaluation de l'apport de la présence d'un ACA à l'attrait et à l'efficacité d'une aide en ligne à l'utilisation de Flash, logiciel grand public de création d'animations sur Internet. L'ACA est une tête parlante développée par France Télécom R&D, le système d'aide multimodal (parole + graphique) réalisé à l'aide de notre plate-forme est destiné à familiariser des utilisateurs novices à l'utilisation de Flash.

### 3.1. Poste « utilisateur »

Pour consulter l'aide en ligne, les participants utilisent uniquement la souris. Un panneau de boutons leur permet d'indiquer le type de leur question (demande d'informations procédurales ou fonctionnelles, demande d'explications ou de confirmation) ; ils peuvent préciser l'objet sur lequel porte la question en sélectionnant soit un mot dans un lexique hiérarchique affiché dans la fenêtre d'aide, soit un « widget » dans la fenêtre principale de Flash. Voir la figure 2.

L'enregistrement de la trace de leurs interactions avec Flash et le système d'aide comprend, outre les événements utilisateur et système, les copies d'écran correspondant à ces événements et les coordonnées des mouvements oculaires échantillonnés à 60 Hz.

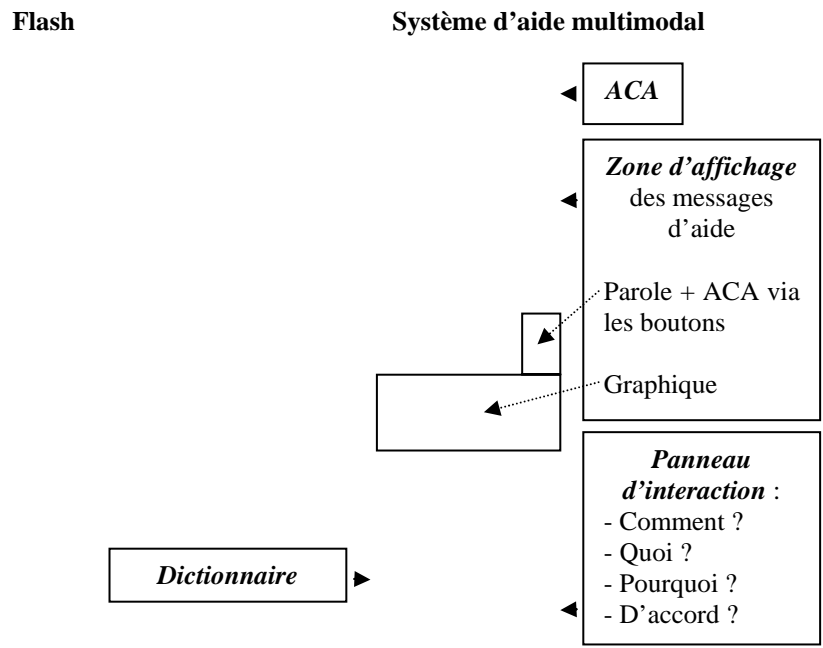

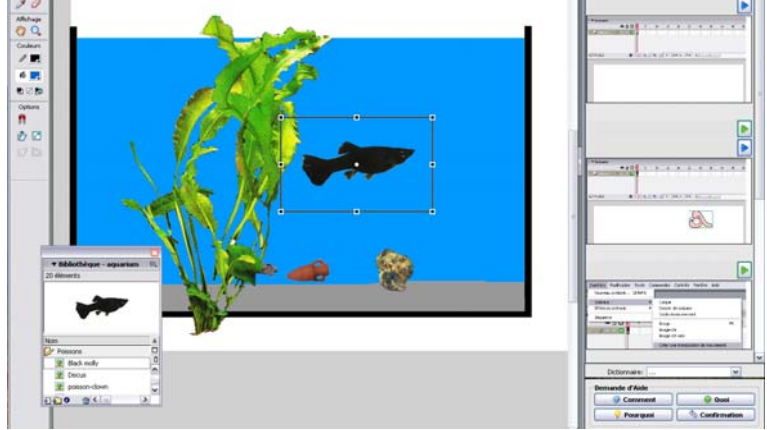

**Figure 2.** *Poste « utilisateur » avec, à gauche Flash, à droite le dispositif d'aide.*

**3.2. Poste « compère »**

L'assistance logicielle dont dispose le compère lui permet, d'une part, d'observer l'activité du sujet à l'insu de celui-ci, d'autre part, de simuler le système d'aide de façon crédible. Plus précisément, la plate-forme fournit au compère :
- L'affichage en temps réel, dans une fenêtre dédiée, du contenu de l'écran du sujet, y compris la position courante du curseur souris.
- La possibilité d'agir sur le logiciel Flash pour annuler la ou les dernière(s) action(s) effectuée(s) par le sujet.
- L'affichage, dans une autre fenêtre, des titres des messages associés à la requête courante du novice.

Les messages sont filtrés en fonction du type de la requête et de l'objet sur lequel elle porte, et suggère une réponse ; le compère n'a plus qu'à valider ce choix ou à sélectionner un autre message qu'il juge plus approprié compte tenu du contexte d'interaction. Il dispose également de quelques messages généraux qu'il peut prendre l'initiative d'envoyer, par exemple lorsque l'utilisateur est bloqué.

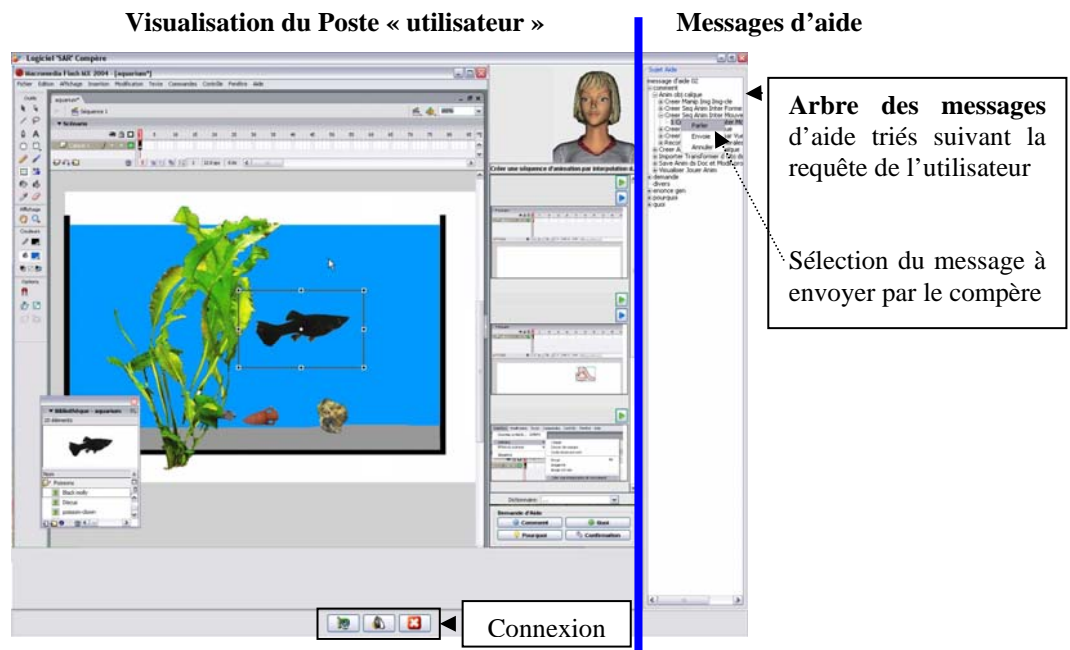

**Figure 3.** *Poste « compère » avec, à gauche, la visualisation de l'affichage de l'utilisateur et, à droite, la fenêtre de sélection/activation des messages d'aide.*

## 4. Conclusion

La plate-forme logicielle d'assistance à l'étude expérimentale de l'interaction homme-machine que nous avons réalisée possède un domaine d'utilisation relativement vaste. En effet, elle permet de simuler avec réalisme, pour un logiciel quelconque, une interface utilisateur multimodale complexe, intégrant éventuellement un ACA, d'enregistrer et de « rejouer » une trace des interactions d'utilisateurs potentiels avec l'interface simulée, comprenant événements utilisateur et système, copies d'écran, énoncés et fixations oculaires de l'utilisateur. L'utilisation actuelle de cette plate-forme (16 sujets sur 20 pendant 1 heure environ) démontre sa robustesse.

## 5. Bibliographie